# Non-Boolean hidden variable model reproduces Quantum Mechanics' predictions for Bell's experiment.

**Alejandro A. Hnilo**

*CEILAP, Centro de Investigaciones en Láseres y Aplicaciones, (CITEDEF-CONICET);*
*J.B. de La Salle 4397, (1603) Villa Martelli, Argentina.*
*email: alex.hnilo@gmail.com*

*Abstract.*

The experimentally verified violation of Bell's inequalities apparently implies that at least one of two intuitive beliefs, which are accepted as true in almost all scientific practice, must be false: that effects propagating at infinite velocity do not exist, and that natural phenomena occur independently of being observed. Giving up any one of these two beliefs (usually known together as Local Realism) is controversial. Many theories have been proposed to reconcile the observed violation of Bell's inequalities with Local Realism, but none has been fully successful. In this paper, it is recalled that Bell's inequalities are equivalent to the conditions to decide the completeness of a theory according to Boolean logic. Therefore, any theory aimed to violate Bell's inequalities must start by giving up Boolean logic. The problem is hence split in two: the "soft" problem is to explain the violation of Bell's inequalities without violating (non-Boolean) Local Realism. The "hard" problem is to predict the time values when single particles are detected, in such a way that the resulting number of coincidences violates Bell's inequalities. A simple hidden variables model is introduced, which solves the "soft" problem in an even ideally perfect setup. This is possible thanks to the use of vectors in real space as the hidden variables and the corresponding operation (projection), which do not hold to Boolean logic. This model reconciles the violation of Bell's inequalities with Local Realism and should end decades of controversy. Regarding the "hard" problem, the introduced model is as incomplete as Quantum Mechanics is. It is argued that solving the "hard" problem involves devising a new kind of quantum computer, which should be able to accept (non-Boolean) hidden variable values as input data and replace the statistical Born's rule of usual Quantum Mechanics with a deterministic threshold condition.

## 1. Introduction.

The "Einstein-Podolsky-Rosen (EPR) paradox" is probably the most discussed problem in the History of Physics. In 1978, C.Cantrell and M.Scully humorously stated there were $\approx 10^6$ papers published about it [1], and each week more papers are produced than any person could ever hope to read. The problem can be briefed as follows: during the early years of Quantum Mechanics (QM), arguments showing the impossibility of observing an electron without introducing disturbances were presented to justify why position and momentum cannot be measured simultaneously with arbitrary precision (the uncertainty principle), and hence, why predictions in the quantum realm can be only statistical. A.Einstein distrusted this non-deterministic feature, and presented one objection after another. In 1935, EPR observed that those arguments failed to explain the high correlations existing between outcomes of observations performed on distant particles that had interacted in the past. Such correlations must exist because of conservation of total momentum and position of center of mass. EPR noted that it was conceivable measuring momentum in one particle and position in the other with arbitrary precision and in a fully independent way (unless some sort of "spooky action at a distance" existed), circumventing the uncertainty principle. EPR then concluded that some elements were missing in the description of physical reality provided by QM. The hypothetical missing elements, assumed unobservable, were later named hidden variables (HV). Note the subtle difference: EPR objected the completeness of the *description of physical reality*, not the completeness of QM theory. Bohr's not-less-famous answer [2] criticized the idea of physical reality independent of the observer.

In 1965, Bell's inequalities apparently demonstrated the completion asked by EPR to be an impossible task without breaking intuitive ideas about Locality, or Realism [3] (in short: Local Realism). Bell's argument does not deal with the original EPR gedanken experiment, but with one similar to the sketch in Figure 1 (EPR-Böhm experiment). In this setup, the key measured magnitude is the number of *coincidences*, that is, the number of particles detected at the same time in both sides of the experiment, as a function of the angle settings $\{\alpha,\beta\}$. A complete description of the physical reality of this experiment in the EPR-sense must be able to predict the *time values* when particles are detected at each output gate (+1 or -1) in each station, and the resulting number of coincidences must fit QM predictions. Note that QM is not able to predict these time values, but only average rates. F.ex., let call $N^{++}$ the number of simultaneous detections at both gates +1 in Fig.1 (in the same way: $N^{--}$, $N^{+-}$, $N^{-+}$ and $N = \Sigma\, N^{ij}$). For the fully symmetrical, maximally entangled state $|\varphi^{+}{}_{AB}\rangle = (1/\sqrt{2}).\{|x_A,x_B\rangle + |y_A,y_B\rangle\}$ in the usual notation, QM predicts (for a sufficiently long recording time):

$$N^{++}/N = N^{--}/N = \tfrac{1}{2}.cos^2(\alpha-\beta);\; N^{+-}/N = N^{-+}/N = \tfrac{1}{2}.sin^2(\alpha-\beta) \qquad (1)$$

which violates Bell's inequalities.

The verification of the violation of Bell's inequalities in nearly ideal (or *loophole-free*) conditions [4-7] leads many researchers to declare (concurring with Bohr) that physical reality independent of the observer is inexistent. The issue is important, because the existence of such reality is assumed not only in everyday's life, but also in all scientific practice (excepting QM and, perhaps, some Humanities). The alternative is accepting the existence of non-local effects, but it risks conflict with the theory of Relativity. The tension between the two fundamental theories has been described by stating that QM and Relativity are in a "peaceful coexistence".

Hoping to reconcile QM with Local Realism, many HV theories have been proposed for decades to provide a complete description, in the EPR-sense, of the physical reality of Fig.1 experiment. But, none has been able to achieve this goal in a fully satisfactory way. None (nor QM) is able to predict the time values detections occur.

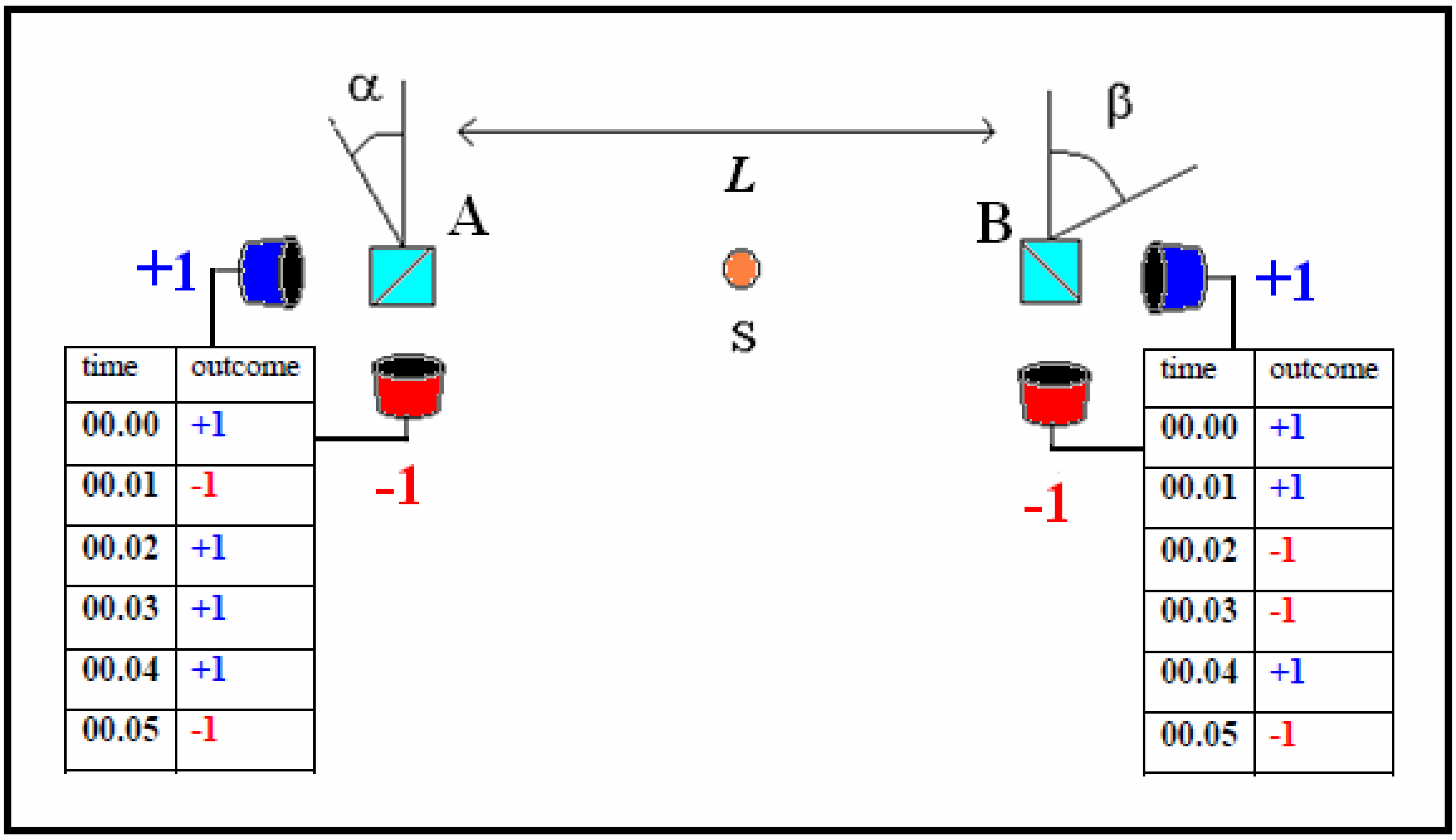

Figure 1: Sketch of an ideal Bell's (or EPR-Böhm's) experiment. Source **S** emits states entangled in polarization, which propagate to stations **A** (observer: Alice) and **B** (Bob) separated by a (large) distance ***L***. Single particles are detected after analyzers set at angles {α,β}. Each station has a device that stores the time values particles are detected. F.ex., at time 00.02, detections occur at gates +1 in **A** and -1 in **B**, what is a (+1,-1) coincidence. The number and type of coincidences as a function of {α,β} violates classical bounds (Bell's inequalities). The complete description of the physical reality of this experiment in the EPR-sense must be able to predict the time values particles are detected at each gate, and to fit QM predictions.

In this paper, the issue is split in two: the first ("soft") problem is to solve the apparent contradiction between the violation of Bell's inequalities and Local Realism. The second ("hard") one is to predict the time values when particles are detected (within Local Realism, and fitting eq.1) at each gate. Solving the second problem also solves the first one. Most HV theories have attempted to solve the issue in a single strike by solving the second problem. The intrinsic difficulties involved (that are discussed later) are probably the cause why no attempt has reached full success yet. The first problem instead, once the necessity of *non-Boolean* logic is recognized, has an almost trivial solution. In short: the soft problem is really soft, and the hard one is fundamentally hard.

A simple HV model is introduced that solves the soft problem. It reproduces QM predictions for Bell's experiment in ideal conditions without violating (non-Boolean) Local Realism. A solution to the hard problem is suggested, which involves the development of a new type of quantum computer (which here means a logical framework, rather than a device). In the next Section 2, the relationship between two different definitions of Locality and Realism, and Boolean logic, is reviewed. In Section 3, two HV models are contrasted: a Boolean one (which holds to Bell's inequalities), and a vector-based, non-Boolean one (which violates them). In Section 4 some relevant questions are discussed.

Before going on, it is pertinent to note that the expression "quantum non-Locality" is often used to mean "quantum, as opposed to classical". However, arguments following the Copenhagen interpretation [8-11] demonstrate the violation of Bell's inequalities to be the consequence of the uncertainty principle applied to *local* variables. In other words: of the wave nature of each single quantum system. As stated in Reference 8, page 184: *"Violations of Bell's inequality…are due to the wave properties of matter...a HV approach which explicitly takes into account the wave properties of matter may lead to results similar to those of QM…"*. The HV model introduced in this paper can be read as one of the possible realizations of that approach.

## 2. Locality, Realism and a simple non-Boolean HV model.

### *2.1 Boolean and non-Boolean Local Realism.*

Actually, it has been known for a long time that QM predictions are compatible with *some* forms of Local Realism. A paper published in 2020 cited more than 70 articles pointing out that [12]. Yet, despite some of these articles are more than 30 years old, the controversy QM vs. Local Realism remains. The reason of the controversy's survival is, perhaps, the lack of a simple example showing that compatibility. Hopefully, the example introduced in this paper will fill that gap.

Much of the problem (and the solution) lays in the definition of "Locality" and "Realism". The complete issue is complex and subtle [13,14]. For the purposes of this paper, let condense it into two options:

Realism #1: the outcome of the measurement of any physical magnitude is definite regardless the measurement is actually performed and observed, or not (roughly speaking: *counterfactual definiteness*).

Realism #2: the same as #1 and besides, the *probability* of occurrence (including 0 and 1) of a given outcome can be calculated as a (Riemann or Lebesgue) integral on a probability distribution $\rho(\lambda)$ and an observation probability $P(\lambda,j)$, where $\lambda$ is a (possibly unobservable, or hidden) variable and $j$ are parameters controllable by the observer; $\rho(\lambda)$ and $P(\lambda,j)$ hold to classical (Kolmogorov's) axioms of theory of probability.

Locality #1: the measurement's outcome is independent of anything outside the past half light-cone of said measurement (roughly speaking: *Einstein's Locality*).

Locality #2: the same as #1 and besides, the probability of simultaneous outcomes $a$ and $b$ ($a$, $b$ = +1 or -1 in Fig.1) of two measurements performed at remote stations A and B, with parameters respectively $\alpha$ and $\beta$, holds to:

$$P_{AB}{}^{ab}(\lambda,\alpha,\beta) = P_A{}^a(\lambda,\alpha).P_B{}^b(\lambda,\beta) \tag{2}$$

From definitions #2 it follows that the probability of observing outcomes $a$ and $b$ is:

$$P_{AB}{}^{ab}(\alpha,\beta) = \int d\lambda\,.\rho(\lambda).P_A{}^a(\lambda,\alpha).P_B{}^b(\lambda,\beta) \tag{3}$$

which is essential to derive Bell's inequalities [3]. Besides, in order to derive Bell's inequalities, the distribution $\rho(\lambda)$ must *not* be a function of $\alpha,\beta$ (*non-contextuality*, see also Section 4.2 later). Assuming contextuality valid apparently means that the source "knows" the values $\alpha,\beta$ a time (at least) $L/2c$ in advance, despite $\alpha,\beta$ can be changed at will by the observer. It seems to imply an effect that propagates backwards in time. Non-validity of eq.2 and contextuality are usual ways to introduce non-local effects to explain the observed violation of Bell's inequalities.

The difference between definitions #1 and #2 is the use of (classical) probabilities. Classical probabilities presuppose Boolean logic [15]. Boolean logic is familiar: it states that something has a certain feature, or it has not. It deals with elements belonging to measurable sets, and sets included (or not) in other sets. By adding the operations of union ($\cup$) and intersection ($\cap$) of sets, a Boolean algebra is defined. Boolean logic and algebra are well known, they can be visualized with Venn's diagrams. J.Bell naturally assumed classical probabilities (and hence, Boolean logic) to describe the experiment in Fig.1. However, Boolean logic implies Boole's conditions of completeness. G.Boole devised them, more than one century ago, as a criterion to decide whether a theory (any theory) can be considered *complete*, or not. Boole's conditions, when applied to the experiment in Fig.1, are equivalent to Bell's inequalities [16,17]. In consequence, violating Bell's (or Boole's) inequalities within a Boolean theory (which includes classical probabilities) is a logical impossibility. A HV theory with some hope to violate Bell's inequalities (in general: to fit QM predictions) must start by giving up Boolean logic. Indeed, QM is non-Boolean.

### *2.2 A non-Boolean HV model using vectors.*

In this paper, the general term "filtering" means selecting systems having more than one attribute. In Boolean logic, systems having a certain attribute are represented by a *set*. Systems having more than one attribute are then found by *intersection* of the sets corresponding to these attributes. A simple non-Boolean description involves *vectors* and vectors' *projection* as a related operation. Filtering now means projection of the corresponding vector. F.ex.: exclusive attributes, which are represented by disjoint sets in Boolean logic, are now represented by orthogonal vectors.

In what follows, vectors are indicated with bold typing, scalars with normal typing. The operation projection of vector **b** on vector **a** is:

$$\mathbf{a}.\mathbf{b} \equiv \mathrm{b}.cos(\gamma).\mathbf{e}_a = (\mathbf{a}\times\mathbf{b}/\mathrm{a}).\mathbf{e}_a \qquad (4)$$

where "**.**" means projection, "." means multiplication with a scalar, "×" means scalar product of vectors, γ is the angle between **a** and **b**, b (a) is the modulus of **b** (**a**), and $\mathbf{e}_a$ is the unit vector in the direction of **a**. The projected vector represents the systems having attributes **b** and (then) **a**. Projection is neither commutative nor associative. It implies a non-Boolean algebra.

Replacing sets by vectors is more disruptive than it may seem. Consider, for example, systems having exclusive attributes A and B. In the Boolean case, A∩B = 0 no matter what other operations or filtering are performed. In the vectors' case, **A.B** = 0 but **A.C.B** ≠ 0 if **C** is oriented at an intermediate angle between **A** and **B**. This is the well known result for a polarizer inserted at an intermediate angle between two crossed polarizers, but it is a weird result from the point of view of Boolean logic. Imagine "A" is the attribute of being a sphere, and "B" being a cube. Then imagine that other attributes are being blue (attribute "C") or red. Then **A.C.B** ≠ 0 means that one can start with physical systems having the attribute of being *cubes* (red and blue) and, after filtering the ones having the attribute of being blue, get *spheres* at the end.

Eq.2 looks much like the projection of state |b⟩ into state |a⟩ in QM: |a⟩⟨a|b⟩. Yet, the elements of QM algebra are not simple vectors, but closed vector subspaces or their corresponding orthogonal manifolds in a Hilbert's space. QM algebra is an ortho-complemented non-distributive lattice [18] and is (of course) non-Boolean. Nevertheless, vectors in real space suffice to understand typical quantum effects as interference, superposition, non-commutative operations (rotations) and, as it will be shown, the violation of Bell's inequalities. As it has been already noted, the main problem is how to link vectors' length, which is a continuous variable, to discontinuous particle detection. A.Khrennikov has named this to be "the true quantum problem". Born's rule of QM is the simplest solution: the normalized vector state's squared modulus gives the *probability* of detecting a particle. But, this solution fatally limits QM to be a statistical description. The next simplest solution is to relate particle detection with some *threshold* condition [19].

Suppose then that the physical system under consideration carries a vector HV named **V**(t):

$$\mathbf{V}(\mathrm{t}) = f(\mathrm{t}).\mathbf{e}_x + g(\mathrm{t}).\mathbf{e}_y = \mathrm{V}(\mathrm{t}).\mathbf{v}(\mathrm{t}) \qquad (5)$$

which varies arbitrarily in time. Here $\mathbf{e}_x$, $\mathbf{e}_y$ are unit vectors in the plane perpendicular to the system's direction of propagation, and $f$(t) and $g$(t) are arbitrary functions. The modulus of **V**(t) is V(t); the unit vector in its direction is **v**(t), which is at an angle ν(t) with the *x*-axis. For non-polarized systems V(t) and ν(t) are statistically independent. It is assumed that in a (long) experimental run of duration *Tr*, the number *N* of detected particles is:

$$N = (1/u).\int_0^{Tr} dt.\,|\mathbf{V}(t)|^2,\ N{>>}1 \text{ assumed} \qquad (6)$$

where *u* is a threshold value. Be aware that *N*>>1 is assumed. Therefore, the predictions of this model are valid in the statistical limit only (as the QM ones are).

It is tempting, and perhaps helpful, thinking **V**(t) as a physical field, say, the electric field. But it must be kept in mind that **V**(t) is a hypothetical HV. It can have the features that are found convenient as far as no contradiction arises. As all HV models, this one is neither intended to replace QM nor is claimed to be an accurate description of physical reality. But, as it will be shown, it makes clear why there is no contradiction between the violation of Bell's inequalities and Locality or Realism (definitions #1, of course, for definitions #2 presuppose Boolean logic and are thus, from start, incompatible with that violation).

Systems that pass through the +1 gate of a polarization analyzer oriented at an angle α (in other words: that are filtered by the analyzer) are represented by the vector component obtained by projection of **V**(t) into the direction $\mathbf{e}_\alpha$ (this statement can be read as the *definition* of a polarization analyzer). The detected number of particles $N_\alpha$ at that gate of the analyzer is then:

$$N_\alpha = (1/u).\int_0^{Tr} dt.\,|\mathbf{e}_\alpha.\mathbf{v}(t).V(t)|^2 = (1/u).\int_0^{Tr} dt.\,|\mathbf{e}_\alpha.cos[\nu(t)-\alpha].V(t)|^2 = (1/u).\int_0^{Tr} dt.\,V^2(t).cos^2[\nu(t)-\alpha] \qquad (7)$$

for non-polarized systems ν(t), V(t) and α (which is chosen by the observer) are independent, hence the average value of the cosinus-squared factor is ½:

$$N_\alpha = ½.N \qquad (8)$$

for all values of α (as expected).

All the necessary elements to explain the violation of Bell's inequalities within Local Realism (definitions #1, of course) are at hand now: the number of coincidences is given by a time integral in which the integrand is the squared modulus of a vector. This vector is the one that represents the systems having the features of being transmitted through analyzers oriented as $\mathbf{e}_\alpha$ and $\mathbf{e}_\beta$. Therefore, it is found by projection of **V**(t) into $\mathbf{e}_\alpha$ and $\mathbf{e}_\beta$ (or vice versa). The resulting factor $\mathbf{e}_\beta.\mathbf{e}_\alpha$ (or $\mathbf{e}_\alpha.\mathbf{e}_\beta$) is independent of time, and produces the sinusoidal curve that violates Bell's inequality. The next Section 3 is devoted to explain this reasoning in detail. The Reader who finds such explanation unnecessary may go directly to Section 4, where some questions are discussed.

## 3. Boolean and non-Boolean hidden variable models.

*3.1 A simple Boolean HV model.*

In order to make evident, by comparison, the features of the (non-Boolean) vector-HV model, let first introduce a classical, Boolean HV one [20]. In this model, each single system carries an angular HV named $\lambda \in [0,\pi]$ that varies arbitrarily from one system to the next. Polarization analyzers operate in the following way:

$$\text{If } \lambda \in [\alpha-\pi/4,\ \alpha+\pi/4]\text{, then: } P^+(\lambda,\alpha) = 1 \text{ and } P^-(\lambda,\alpha) = 0, \tag{9}$$

$$\text{If } \lambda \notin [\alpha-\pi/4,\ \alpha+\pi/4]\text{, then: } P^+(\lambda,\alpha) = 0 \text{ and } P^-(\lambda,\alpha) = 1,$$

where $P^+$ ($P^-$) is the probability to be transmitted (deflected) at the polarizer, see Figure 2. As the model is assumed to hold to Boolean logic, a classical probability distribution $\rho(\lambda)$ can be assigned. For a non-polarized incident ensemble, $\rho(\lambda)$ is defined uniform in $[0,\pi]$. The number of particles detected after a polarizer during time *Tr* is then:

$$N_\alpha = \int_0^{Tr} dt.\ \rho[\lambda(t)].P^+(\lambda(t),\alpha) = ½.N \tag{10}$$

for all values of $\alpha$ (as expected).

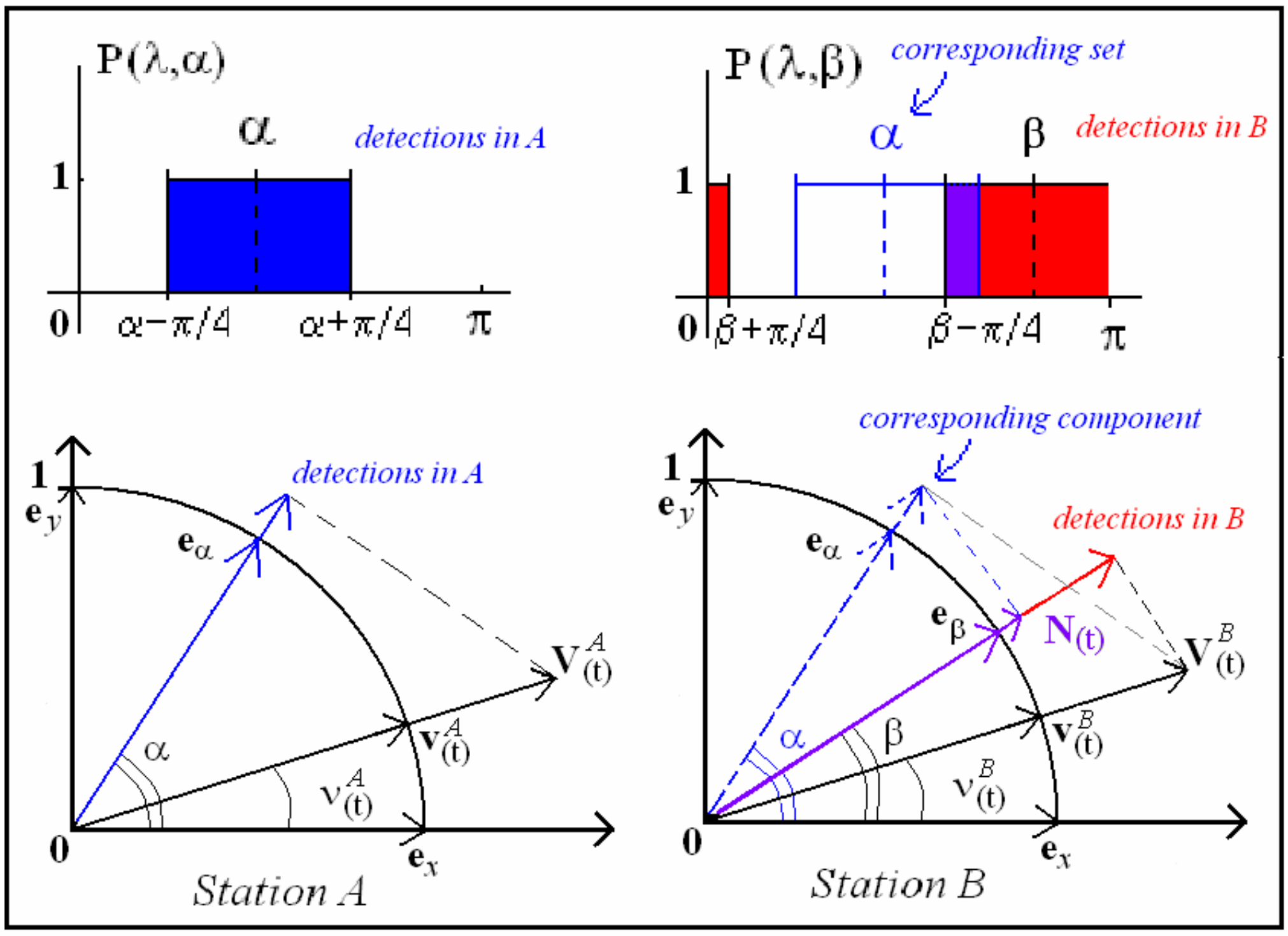

Figure 2: **Up, Boolean model:** In Station A, the set of $\lambda$ that produces detections after the analyzer oriented at $\alpha$ is the interval $[\alpha-\pi/4,\alpha+\pi/4]$ (blue). In station B, it is the interval $[\beta-\pi/4,\beta+\pi/4]$ (red). Coincidences are produced in the intersection (violet) of the "corresponding set" with $[\beta-\pi/4,\beta+\pi/4]$. **Down, vector-HV model:** In Station A, the vector component that produces detections after the analyzer oriented at $\alpha$ is the projection of the vector HV into the direction $\mathbf{e}_\alpha$ ("*detections in A*" vector, blue); in station B, it is the projection into the direction $\mathbf{e}_\beta$ ("*detections in B*" vector, red). Coincidences are produced by the component **N**(t) (violet), which is the projection of the "corresponding component" into the direction $\mathbf{e}_\beta$. Both processes (Boolean and non-Boolean) are strictly local. This figure applies to state $|\varphi^+_{AB}\rangle$.

Let consider now the case of entangled systems. For the symmetry of the Bell state $|\varphi^+_{AB}\rangle$, each system carries a HV with the same value $\lambda^B(t) = \lambda^A(t)$ (for other Bell's states, f.ex. for $|\psi^-_{AB}\rangle$, $\lambda^B = \lambda^A - \pi/2$). The HV are counterfactual definite (Realism #1). Hence, in the space of station B there exist a set of HV (named here "corresponding set") that corresponds to the $\lambda^A(t)$ producing (+1) detections in A, even if no observation is actually performed (Fig.2, upper line). This set has the same size and (for $|\varphi^+_{AB}\rangle$) position than the one producing detections in the space of station A. The existence of the "corresponding set" in B-space does not imply action-at-a-distance. Bob, who is supposed to have access to the values taken by the HV, may think: "If Alice had oriented her analyzer at angle $\alpha$, then this corresponding set would be the right one to calculate coincidences". Bob knows this even if he doesn't know what Alice actually does. In this Boolean model, Bob can calculate not only the number of coincidences $N^{++}$, but he also knows (if he knows the value of $\lambda^B$) *when* a coincidence *would* occur at each gate for each possible choice taken by Alice. The description of physical reality is complete in the EPR-sense, but the number of coincidences does not fit eq.1 (see below).

The $\lambda^B$ that take part in the calculation of $N^{++}$ belong not only to the corresponding set, but also to the "transmitting" set in B-space. The set of HV producing (+1,+1) coincidences is obtained, by local filtering, as the intersection of both sets. Time integration gives the number of coincidences:

$$N^{++} = (1/u) \int_0^{Tr} dt.\, \rho[\lambda(t)].P(\lambda^A(t),\alpha).P(\lambda^B(t), \beta) = [\tfrac{1}{2} + (\beta-\alpha)/\pi].N \qquad (11)$$

If the calculation is performed in the A-space the result is clearly the same, because the operation "intersection" is commutative. Eq.11 does predict exact correlation or anti-correlation (see Figure 3), but it does *not* violate Bell's inequalities (this is known from start, for the model is Boolean).

### *3.2 Entanglement in the vector-HV model.*

In the vector-HV model, if the systems are entangled as the Bell state $|\varphi^+_{AB}\rangle$, then their HV are related as: $\mathbf{V}^B(t) = \mathbf{V}^A(t)$, thus $\nu^B(t) = \nu^A(t)$ (f.ex. for $|\psi^-_{AB}\rangle$ instead, $\nu^B = \nu^A - \pi/2$). In the Boolean model the integral over the appropriate *set* of HV calculates the number of detected particles. In the vector-HV model, it is the integral of the appropriate *vector* HV what allows that calculation. Recall that, because of Realism #1, $\mathbf{V}^A(t)$, $\mathbf{V}^B(t)$ and all their components are definite in each station, regardless a measurement is performed, or not. Note the reasoning is parallel to the one in the Boolean case:

Detections in gate +1 of station A are produced by the vector component which is the projection of $\mathbf{V}^A(t)$ into the axis $\mathbf{e}_\alpha$, or $\mathbf{e}_\alpha.\mathbf{V}^A(t)$. The vector-HV are counterfactual definite (Realism

#1). Hence, in the space of station B there exist a vector (named here "corresponding component") that corresponds to the $\mathbf{e}_\alpha.\mathbf{V}^A(t)$ producing detections in the A-space, even if no observation is actually performed (Fig.2, lower line). This vector has the same size and (for $|\varphi^+_{AB}\rangle$) orientation as the one producing detections in the A-space. As in the Boolean case, the existence of the "corresponding component" (which is a component of $\mathbf{V}^B$) in B-space does not imply action-at-a-distance. Bob, who is supposed to have access to the values taken by the HV, may think: "If Alice had oriented her analyzer at angle α, then this corresponding component would be the appropriate one to calculate coincidences". Bob knows this even if he doesn't know what Alice actually does. In this vector-HV model, he can calculate the number of coincidences $N^{++}$ but he doesn't know *when* a coincidence occurs at each gate. The description of physical reality is *not* complete in the EPR-sense. The "hard" problem is not solved.

The vector component that takes part in the calculation of $N^{++}$ belongs not only to the corresponding component, but also to the "transmitting" component in B-space. Therefore, the component that produces coincidences is obtained, by local filtering, as the *projection* of the "corresponding component" into the transmitting component of B (Fig.2, down, right), i.e., $\mathbf{N}(t) \equiv \mathbf{e}_\beta.\mathbf{e}_\alpha.\mathbf{V}^B(t)$. Time integration gives the number of coincidences:

$$N^{++} = (1/u).\int_0^{Tr} dt.\,|\mathbf{N}(t)|^2 = (1/u).\int_0^{Tr} dt.\,|\mathbf{e}_\beta.\mathbf{e}_\alpha.\mathbf{v}^B(t).V^B(t)|^2 = (1/u).\int_0^{Tr} dt.\,|\mathbf{e}_\beta.\mathbf{e}_\alpha.cos^2[\nu^B(t)\text{-}\alpha].V^B(t)|^2 =$$

$$= cos^2(\alpha\text{-}\beta).(1/u.).\int_0^{Tr} dt.\,cos^2[\nu^B(t)\text{-}\alpha].V^2(t) = cos^2(\alpha\text{-}\beta).½.N, \text{ recall that } N\gg 1 \qquad (12)$$

According to the usual interpretation of probability as the limit of frequencies, then:

$$P^{++} \equiv N^{++} / N = ½.cos^2(\alpha\text{-}\beta) \qquad (13)$$

which is the QM prediction for the state $|\varphi^+_{AB}\rangle$ and violates Bell's inequalities. Despite the operation "projection" is not commutative, the final result is the same regardless the vector component to calculate coincidences is performed in A-space or in B-space. This is because the projected vector's squared modulus is all what accounts to calculate the integral in eq.12. Note that the HV-model does not use probabilities, but a deterministic threshold condition. Yet, as it is valid for $N\gg 1$ only, its predictions are statistical (as the QM ones are).

The probabilities of coincidences involving (-1) gates (i.e.: $P^{-+}$, $P^{+-}$, $P^{--}$) are calculated in the same way, by defining unit vectors $\mathbf{e}_{\beta\perp}$ and $\mathbf{e}_{\alpha\perp}$. The results for the other Bell states are obtained assuming that $\mathbf{V}^B(t)$ is emitted rotated respect to $\mathbf{V}^A(t)$. Assuming $\mathbf{V}^A(t) = f(t).\mathbf{e}^A{}_x + g(t).\mathbf{e}^A{}_y$, the appropriate rotations are:

$$\text{for } |\varphi^-_{AB}\rangle\text{: } \nu^B(t) = -\nu^A(t), \text{ or } \mathbf{V}^B(t) = f(t).\mathbf{e}^B{}_x - g(t).\mathbf{e}^B{}_y \Rightarrow P^{++} = ½.cos^2(\alpha+\beta),$$

$$\text{for } |\psi^-_{AB}\rangle\text{: } \nu^B = \nu^A - \pi/2, \text{ or } \mathbf{V}^B(t) = g(t).\mathbf{e}^B{}_x - f(t).\mathbf{e}^B{}_y \Rightarrow P^{++} = ½.sin^2(\alpha\text{-}\beta), \qquad (14)$$

for $|\psi^{+}{}_{AB}\rangle$: $\nu^B = \pi/2 - \nu^A$, or $\mathbf{V}^B(t) = g(t).\mathbf{e}^B{}_x + f(t).\mathbf{e}^B{}_y \Rightarrow P^{++} = ½.sin^2(\alpha+\beta)$.

### *3.3 Comparison between the two models, Locality.*

The Boolean HV model "solves" the second problem (and hence, also the first one) mentioned in the Introduction, because λ(t) determines not only the number of coincidences observed during a long time, but also the time values when particle detection and coincidences occur at each gate. But, the obtained result (eq.11) fails to reproduce QM predictions and observations. Eq.11 is a saw tooth function which is exactly at the limit allowed by Bell's inequalities (or by Locality and Realism #2). Both QM and the vector-HV model predict eq.12 instead, which violates that limit for (almost) all angles, see Fig.3.

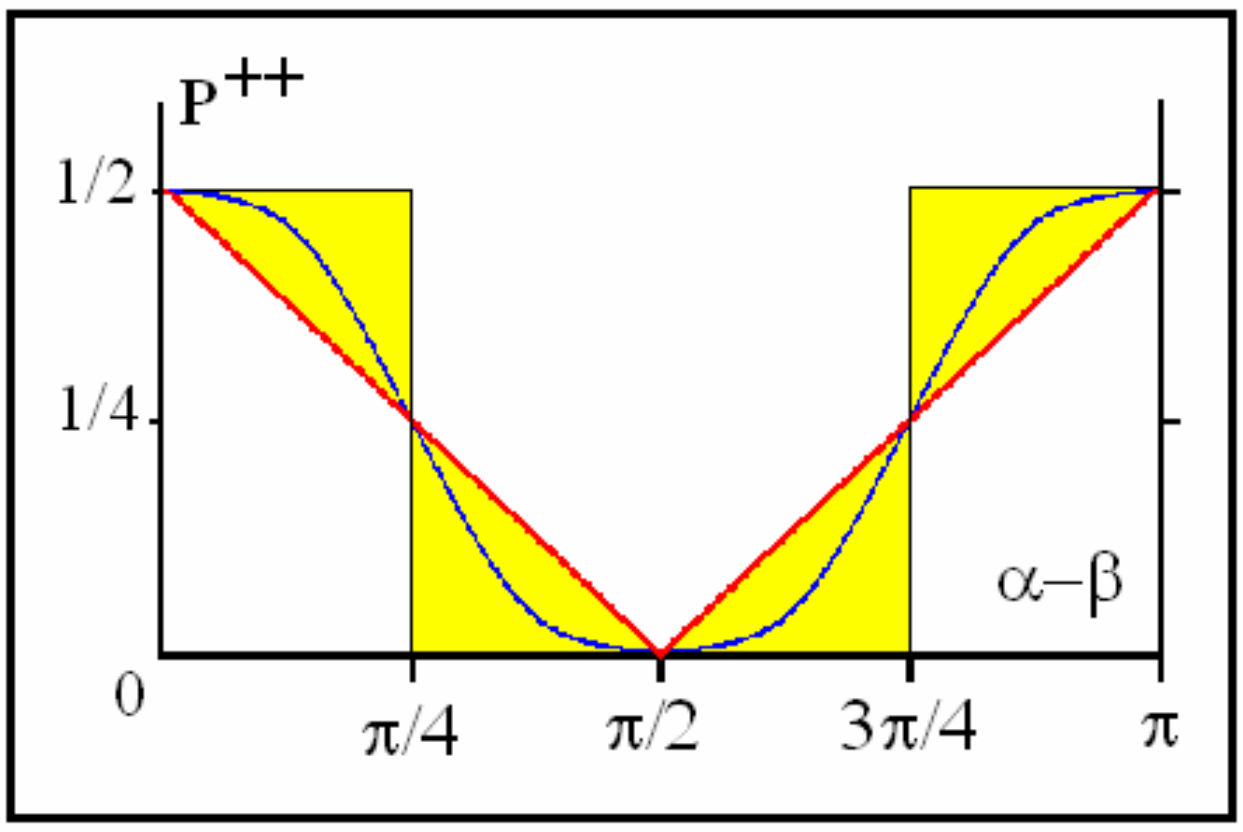

Figure 3: In the Boolean model, the probability of coincidences is given by a saw tooth function (in red for $|\varphi^{+}{}_{AB}\rangle$) which is the limit to classical correlation imposed by Boole-Bell's inequalities. The vector-HV model and QM are non-Boolean and predict a curve (blue) which enters into the "forbidden" region (yellow). Warning: the difference between the red and blue lines is exaggerated to enhance visibility.

It might be thought that some nonlocal effect is lurking in the "corresponding component" in Fig.2. But, note that it does not mean a violation of Locality #1 more than the "corresponding set" in the Boolean model. Both "corresponding things" are the consequences of Realism #1 and the use of HV to create correlations between remote observations. The differences between the Boolean and the non-Boolean models are the nature of the HV (sets and vectors) and the filtering operations involved (intersection and projection).

In the Boolean set-HV model, the systems at stations A and B share a property that is well defined since the moment of their emission: the Boolean HV named λ(t). Detection after an analyzer is determined by the condition that λ(t) belongs to a certain set. The number of coincidences is calculated by local filtering, as the intersection of two sets: the one that corresponds to detection after the analyzer in A with the one that corresponds to detection after the analyzer in B. The operation "intersection" produces a function that varies *linearly* with the angle difference and is exactly in the limit of Bell's inequalities (see Fig.3).

In the non-Boolean vector-HV model, the systems at stations A and B share a property that is well defined since the moment of their emission: the non-Boolean HV named **V**(t). The number of detections after an analyzer is determined by the vector component parallel to the analyzer's axis. The number of coincidences is calculated by local filtering, as the time integral of the squared modulus of a double projection: first of **V**(t) into the axis $\mathbf{e}_\alpha$ that corresponds to detection after the analyzer in A, and then into the axis $\mathbf{e}_\beta$ that corresponds to detection after the analyzer in B. The operation "projection" produces a function that varies *quadratically* (at first order) with the angle difference and violates the limit of Bell's inequalities for (almost) all angles (see Fig.3).

In no case Locality #1 is violated. In the non-Boolean case Locality #2 is meaningless, for probabilities cannot be reliably defined outside Boolean logic (nevertheless, see Section 4.2 below). In both Boolean and non-Boolean cases coincidences are defined by local filtering through the devices in both stations. The difference is that *filtering* is an operation with different meaning (and features) in each case. The two HV models are compared side-by-side in Table 1.

| | Boolean | Non-Boolean |
|---|---|---|
| Hidden variable | $\lambda(t)$ | $\mathbf{V}(t)$ |
| Distribution | $\rho(\lambda)$ | it doesn't exist |
| Detection of $N$ particles | $N = \int dt.\rho[\lambda(t)]$ | $N = (1/u).\int dt.\lvert\mathbf{V}(t)\rvert^2$, $N >>1$ |
| Probability of transmission through an analyzer | $P_A^+(\alpha,\lambda)$, (see eq.7) | it doesn't exist |
| Number of detected particles after an analyzer | $N_\alpha = \int dt.\rho[\lambda(t)].P_A^+(\lambda(t),\alpha) = ½.N$ | $N_\alpha = (1/u).\int dt.\lvert\mathbf{e}_\alpha.\mathbf{V}(t)\rvert^2 = ½.N$ |
| Detection after two analyzers given by | $P_A^+(\alpha,\lambda).P_B^+(\beta,\lambda)$, or integral over $A\cap B$ in $\lambda$-space. | $\mathbf{e}_\beta.\mathbf{e}_\alpha.\mathbf{V}(t)$ |
| Number of coincidences, (calculus) | $N^{++} = \int dt.\rho[\lambda(t)].P_A^+(\alpha,\lambda).P_B^+(\beta,\lambda)$ | $N^{++} = \int dt.\lvert\mathbf{e}_\beta.\mathbf{e}_\alpha.\mathbf{V}(t)\rvert^2$ |
| Number of coincidences, (final expression, see Fig.4) | $N^{++} = [½ + (\beta-\alpha)/\pi].N$ | $N^{++} = ½.cos^2(\beta-\alpha).N$ |

Table 1: Summary of features of Boolean and non-Boolean HV models.

The widespread idea of "quantum non-Locality" arises from giving up eq.2 in order to violate Bell's inequalities. It is meaningful if classical probability, i.e. Boolean logic, is assumed valid. But QM is not Boolean. Therefore, in order to mean "quantum, as opposed to classical", I find the expression "non-Boolean" (or "vector-like", or "wavy", see Section 4.3 below) far more appropriate than "non-Local".

## 4. Some questions.

### *4.1 Isn't the vector-HV model just QM is disguise?*

The Boolean $\lambda(t)$ is like a "bit": it can be only inside (say, $P^+(\lambda) = 1$) or outside ($P^+(\lambda) = 0$) a certain set. Instead, the non-Boolean **V**(t) is like a "qubit": it can be oriented at any superposition of $\mathbf{e}_x,\mathbf{e}_y$. Also, note that it is possible to apply Born's rule to the modulus of **N**(t) and get (after appropriate normalization) the right result for $P^{++}$. It may then be thought that the vector-HV model is just QM in disguise, but important differences do exist.

In order to calculate $P^{++}$ using Born's rule, QM must define the entangled state of two qubits as a single object (f.ex.) $|\varphi^+{}_{AB}\rangle$, which exists in a four-dimension *abstract* space. The object $|\varphi^+{}_{AB}\rangle$ has no internal parts, yet it is spread in *real* space. It is an "atom" (in the original ancient Greek meaning, "it cannot be cut") of arbitrary size. It is not surprising that this object may create the illusion of non-local effects. In the vector-HV model instead, the entangled state is described by two different objects (vectors) existing in real space. They are created with a definite relationship between them at the moment and place of their emission. Afterwards the two objects are fully separated, counterfactual definite, and evolve and act locally. The number of coincidences $N^{++}$ is calculated (in a counterfactual definite way) as the time integral of a vector **N**(t). The vector-HV model shows that eq.13 is compatible with Locality and Realism (definitions #1, of course), and solves the first or "soft" problem. The QM description, instead, is not transparent enough regarding this compatibility; decades of controversy prove this statement.

*4.2 Is it always impossible applying probabilities to non-Boolean problems? Contextuality.*

"Non-Booleanity" does not imply that a description in terms of classical probabilities is always impossible. It just means that using classical probabilities cannot be taken for granted. The Bell's experiment is one example where their use leads to paradoxical results (i.e., apparent violation of Local Realism). Vorob'yev's theorem [21] demonstrates that the description of non-Boolean problems with Boolean tools can be achieved by using *contextual* scenarios. A few words on the meaning of contextuality are in order.

The elementary definition of contextuality was mentioned near eq.2, i.e., a probability distribution which is a function of the settings, say $\rho(\lambda,\alpha,\beta)$ in Fig.1. In a deeper approach [22], the space $\Omega$ of the HV is assumed a classical probability space. Averages over it allow calculating observable probabilities (eq.3). Contextuality then means that it is necessary to define a Boolean sub-algebra $\Omega_B$ for each physical quantity to be observed. These sub-algebras define the different contexts or, in practice, experimental settings (the values of $\alpha,\beta$ in Fig.1).

The origin of contextuality can be visualized as follows. One may think that the angle variable $\nu$(t) of a non-polarized vector **V**(t) has a uniform distribution, see Figure 4a. Yet, as a consequence of the principle of superposition (a vectors', non-Boolean feature) the distribution of $\nu$(t) can also be thought as two sorts of Dirac's deltas centered at $\pi/2$ of each other, see Fig.4b. Which one of these distributions is the correct one? The uniform one or the two deltas? (and in the latter case, which one of the infinite possible pairs?). In the Boolean way of thinking, only an infinitesimal fraction of the uniformly distributed values of ν(t) is parallel or orthogonal to a certain angle α. Yet, these fractions turn out to be here as large as ½, for *any* value of α. This has the flavor of an inversion of cause and effect, a perplexity often found in QM. But there is no such inversion. The correct answer is: no probability distribution is correct. The definition of (classical) probability requires Boolean logic, and vectors are non-Boolean. A probability distribution cannot be properly assigned because of the non-Boolean superposition principle, i.e., the possibility to project **V**(t) into arbitrary axes.

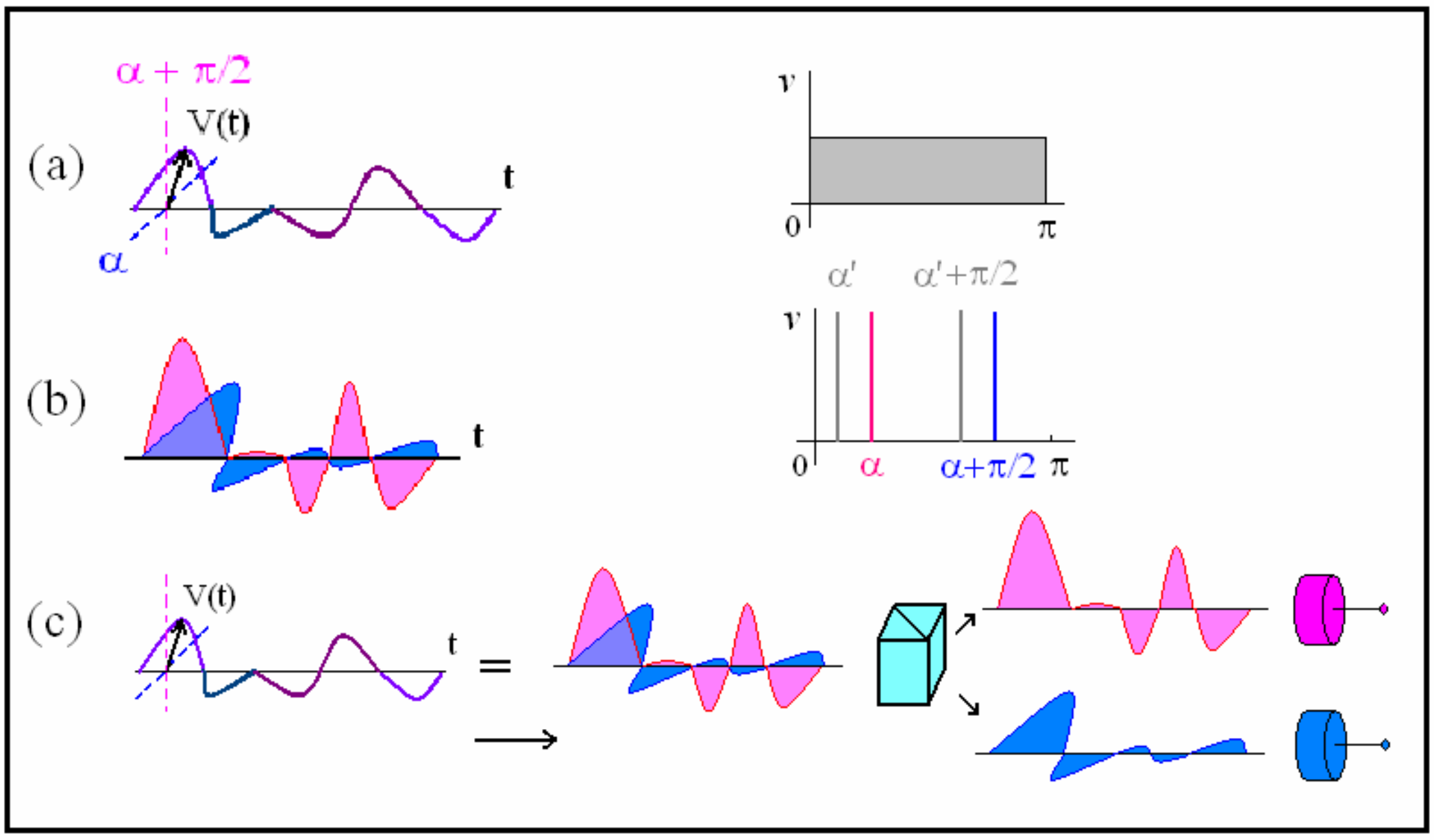

Figure 4: **(a)** *Left:* vector **V**(t) evolves inside and outside the plane of the paper (red in the plane, blue orthogonal to it, different degrees of violet mean intermediate planes). *Right:* For a non-polarized **V**(t), the "probability distribution" of ν(t) is intuitively thought uniform in [0,π]. **(b)**: *Left:* **V**(t) can be projected into two components, one orthogonal to an arbitrary angle α (the plane of the paper) the other one parallel to it. *Right:* as a consequence, the "probability distribution" of ν(t) changes from uniform to two "deltas", placed at α and α+π/2. Besides, α' and α'+π/2 are also possible. This does not mean an inversion of cause and effect, but that a description in terms of probability distributions (which presupposes Boolean logic) is impossible for **V**(t). **(c)**: The vector components that determine whether one detection in each output gate of an analyzer occurs, or not, are well defined *before* the arrival to the analyzer. The analyzer merely separates two already existing components.

When a particle is detected at the +1(-1) gate of an analyzer, it is because the component of **V**(t) parallel (orthogonal) to α reaches some condition for particle detection. This is obviously true after the analyzer; but it is also true *before* the analyzer. If this idea sounds strange, it is because one is used to think in QM terms. In QM, it is stressed that polarization is defined only after the analyzer. This is true within QM theory, but the idea should not sound strange at all for a vector in real space, for an ideal analyzer does not interchange energy with the incident vector. It does not modify nor scramble its components, it just splits them (Fig.4c). The idea simply means that the vector's component parallel (orthogonal) to α reaches the condition to produce detections at gate +1 (-1) even before the analyzer. This condition depends only on **V**(t), α and the detector, and is reached (or not) even if there is no analyzer in place at all. Be aware that only the superposition principle, Realism #1 and an ideal analyzer are involved in this reasoning.

Figure 4 hopefully illustrates the often stated quantum principle: *the observed properties of a physical system depend on the definition of the apparatus of observation* (in Fig.4, the value of the analyzer orientation α). The reason is evident in the vector's realm, and it does not imply violation of Realism (#1, of course). It means that, if a description within Boolean logic is attempted, then the value of α must be defined, i.e. the context (i.e., the $\Omega_B$ sub-algebras mentioned above).

In my opinion, contextuality is a meaningful term within Boolean logic only. If Boolean logic is not valid (as in the vector-HV model, or in QM), then no classical probability space, no Boolean sub-algebra, no probability distribution can be defined, and hence it is not possible to speak of contextuality or non-contextuality. Saying that the vector-HV model and QM are contextual is misleading. They are neither contextual nor non-contextual, they are non-Boolean.

Another way to retrieve a Boolean framework to describe non-Boolean problems is to accept *extended* probabilities, i.e., probability values outside the interval [0,1] [23]. This alternative is well known in Quantum Optics, where a Wigner or a P-Sudarshan distribution function taking negative values is a customary indication of non-classical features. Not surprisingly, extended probabilities are related with contextuality. The difference between the total sum of the probability's modulus and 1 (which is, of course, the total sum in classical probability) defines a *coefficient of contextuality* [24,25]. Other alternative "quasi-Boolean" descriptions of Bell's experiment involve probabilities defined over singular measures [26] or non-measurable sets [27], or p-adic ultrametric distances [28-30] (recall Realism #2).

*4.3 How to complete the description of physical reality in the EPR-sense?*

The vector-HV is able to reproduce QM predictions (including their statistical feature) for the Bell's experiment within Locality and Realism (definitions #1, of course). Yet, it leaves a feeling of frustration. This is because its description of physical reality remains as incomplete, in the EPR-sense, as the one provided by QM. QM uses Born's rule to predict probabilities so that it gives up, from start, any hope of description at the single detection level. The vector-HV model is restricted to $N$>>1, hence it is also unable to provide a description at the single detection level. Both QM and the vector-HV model are unable to predict *when* a particle is detected at each gate and, as a result, the number of coincidences. Both leave the "hard" problem unsolved. The question: "how does Nature do it?" remains unanswered.

As said, Vorob'yev's theorem demonstrates that, in the general case, non-Boolean problems can be described by a classical, Boolean computer only if a context is defined. "Computer" here means a logical framework rather than a device. A quantum computer (which is non-Boolean) is able to reproduce QM predictions for the Bell's experiment without that limitation. The circuit or code is one of the simplest of all. It involves two incoming qubits, one Hadamard operator, one CNOT gate, and rotations on each outgoing qubit to set α,β [31]. Born's rule relates the final state's amplitude with the probability of detecting singles and coincidences. This example hopefully illustrates the deep differences between classical and quantum computers. The latter may well be called "wave" computers [32].

Quantum computers (as they are currently known) cannot be loaded with **V**(t) as input data, and Born's rule limits them to provide statistical predictions only. In consequence, although they are non-Boolean, current quantum computers are unable to predict *when* a particle is detected at each gate and solve the "hard" problem. Conceivably, a different sort of quantum (or "wave") computer, still to be devised, might be able to do the task. It should start by giving up Born's rule to link wave amplitude to particle detection. As already noted, a threshold condition is the natural choice. It should also accept non-Boolean HV as input data. The difficulties in developing such "wave" computer (in the first place: that its necessity had not been realized) probably explain why the hard problem has not been solved yet. Anyway, as the values taken by **V**(t) are unknown by hypothesis, quantum computers (as they are currently known) suffice for all tasks of practical interest. A "wave" computer able to solve the hard problem and explain "how Nature does it" is useful for academic research only. Nevertheless, the demonstration of its existence would have consequences in the area of quantum certified randomness. It has been shown that the series of outcomes in the setup of Fig.1 are incomputable (by a classical Turing machine) [33]. If these series could be computed by a "wave" computer, then they would not be necessarily random. They would be random, or not, depending on the features of **V**(t).

Finally, a classical computer can *mimic* the operation of a "wave" computer but, in accordance to Vorob'yev's theorem, it must use contextual instructions. For an example, see [34].

### *4.4 Some remarks.*

The vector-HV model has the following distinct features:

*i*) Probabilities are not involved. Particle detection is determined by a threshold condition (eq.6).

*ii*) Yet, the vector-HV applies to large ($N$>>1) numbers of detected particles only.

*iii*) As a consequence of (*ii*), the nature of the predictions is statistical. The vector-HV model goes no further than QM.

*iv*) The vector (non-Boolean) hidden variables $\mathbf{V}^{A}$(t) and $\mathbf{V}^{B}$(t) evolve and act locally, and are counterfactual definite. Hence, the vector-HV model is able to violate Bell's inequalities within Local Realism (definitions #1, of course).

*v)* Both QM and the vector-HV model are unable to solve the hard problem. But, while QM proposes no way towards the solution, the vector-HV model does. This is: to operate the vector-HV in a non-Boolean (or "wave") computer of a new kind.

*vi)* The loophole-free experiments violating Bell's inequalities involved Eberhardt states [4,5] or the phenomena of Hong-Ou-Mandel and entanglement swapping [6,7]. In order to give a complete account of the observed violation of Bell's inequalities, it would be desirable the vector-HV to describe these phenomena. Such description is sketched in the Supplementary Material section.

**5. Summary and comments.**

In this paper, the "EPR paradox" (in other words, the problem of completing the description of physical reality of entangled states provided by QM) is split in two: the first (or "soft") problem is to show that the violation of Bell's inequalities and Local Realism are compatible. This is logically possible for the non-Boolean definitions #1 of Local Realism, but not for the definitions #2 involved in the derivation of Bell's inequalities. Once this limitation is recognized, the solution of the soft problem is simple (i.e., the vector-HV model). The second (or "hard") problem is to present a model (holding to Local Realism #1) able to predict not only statistical detection rates, but also the time values single particles are detected, in such a way that the resultant rates of coincident detections violate Bell's inequalities (in general, fit the QM predictions). Solving the hard problem also solves the soft one. Yet, the hard problem is found to be fundamentally difficult, for it requires a non-Boolean logical framework (i.e., a new type of quantum computer) that is still unknown. This difficulty explains why the complete description of physical reality of entangled states, in the EPR-sense, has not been achieved yet.

M.Scully and M.Zubairy have stated that all quantum optical experiments can be explained, at least in a semi-quantitative way, by a semi-classical theory of light (which assumes quantized matter and classical field) plus vacuum fluctuations [35]. They also pointed out two phenomena that unavoidably required field quantization: "quantum beats" and the violation of Bell's inequalities. The vector-HV model presented here can be read as a lesser chapter of the semi-classical theory, specifically devised to explain the violation of Bell's inequalities. This case is important for it is one of the most abundantly and thoroughly performed experimental tests of QM. But, of course, it is not the whole QM. Whether or not the vector-HV model can be expanded to describe quantum beats, and correlations in more complex entangled states (GHZ, cluster, graphs, etc.), is an appealing field for future theoretical research. The aim of the expansion would *not* be, of course, to replace already existing and successful QM descriptions, but to find the limits of the semi-classical approach. Devising a wave computer to solve the "hard" problem is also appealing.

From the experimental point of view instead, the vector-HV model is irrelevant. Its predictions regarding Bell's experiment are the same than those of QM. Nevertheless, showing that QM predictions for the Bell's experiment can be derived without giving up Locality or Realism (definitions #1, of course) does have importance. It should end decades of discussions. In any case, many (including me) will feel more at ease by being able to save Local Realism, than by being forced to give it up.

It is worth remembering here that QM was built as it is, because it had to describe wave phenomena. The condition of stationary (electron's) waves explained the lack of radiated energy from accelerated electrons in the atom and the discrete pattern of spectral lines. Interference of electrons (and other elementary particles as well) was directly observed. Schrödinger equation is a wave equation. The microscopic world is undoubtedly wavy. That's why QM was named *Wave Mechanics* in old textbooks. In turn, vectors are the simplest tool to describe and understand wave phenomena. Young's two-slit experiment cannot be described with classical probabilities, but is easily understood with vectors. Vectors are non-Boolean, and hence can violate Boole's conditions of completeness (= Bell's inequalities). Development of simple vectors' space eventually leads to the Hilbert's space of QM. However, these somehow soothing arguments do not imply that the common perception of the existence of "mysteries" in QM is misled. For, vectors are strange things, stranger than they seem to be at first sight. Many quantum mysteries are just the ones of vectors, but they are mysterious enough [36]: superposition states and their violation of the yes-no logic, apparent violation of the relationship between cause and effect, paradoxical results when the order of filtering (projection) is changed, non-commutative operations (rotations in more than two dimensions, which is related to the Kochen-Specker theorem), failure of the descriptions with classical probabilities and, as it is shown in this paper, violation of Bell's inequalities.

Years ago, I.Pitowski [17] put forward the following question (that he qualified as "the problem of interpretation"): *"WHY is that microphysical phenomena and classical phenomena differ in the way they do?"* He also asked what kind of answer could qualify as a *reasonable* one. In my opinion, an answer qualifies as "reasonable" when it relates the unknown subject to well-known ideas. The vector-HV model then provides such a reasonable answer: the cause of the difference between quantum and classical phenomena is that the former requires a description in terms of non-Boolean entities (say, vectors), while the latter allows a description in terms of Boolean entities (sets). This qualifies as a reasonable answer because vectors, and sets, are well-known ideas learned in secondary school. It is only to be noted that, perhaps, one may have not noticed how alien to intuition the consequences of vectors' features can be.

In lab's jargon, the word *artifact* names a signal that is seen but is inexistent, caused by inappropriate use of instruments. In this sense, the old contradiction between QM and Local

Realism is an artifact caused by using a Boolean instrument (= classical probabilities) to deal with a non-Boolean problem (= wave phenomena).

**Acknowledgements.**

Many thanks to Mónica Agüero, Donald Graft, Federico Holik and Marcelo Kovalsky for the interest, observations and advices. This material is based upon research supported by, or in part by, the U.S. Office of Naval Research under award number N62909-18-1-2021. Also, by grant PIP 2017-027 CONICET (Argentina).

**References.**

## Supplementary Material.

The next Sections sketch a description, in terms of the vector-HV model, of the phenomena involved in the performed loophole-free observations of the violation of Bell's inequalities. This description is more intricate than the one provided by QM. Replacing QM is not the purpose here. The purpose is just to show that the violation of Bell's inequalities can be fully explained within Local Realism (#1, of course). Note that the description is valid even for ideally perfect setups.

### *I. Vector-Bell states. Normalization and orthogonality.*

In QM, the ket $|\phi\rangle$ is an element in a (complex) vector space, and the bra $\langle\phi|$ denotes a linear functional, or an element in the dual space. In the vector-HV model instead, states are represented by simple vectors in real space. In order to describe the phenomena in the next Section, it is convenient to define *Vector-Bell* states (indicated with bold typed letters in what follows) that are analogous to the Bell states of QM. They are obtained from applying the relationships in eq.14 in the main text in a straightforward way:

$$\boldsymbol{\psi_{AB}}^{-}(\mathrm{t}) \equiv \mathbf{V}^{A}(\mathrm{t})\otimes\mathbf{V}^{B}(\mathrm{t}) = [f(\mathrm{t}).\mathbf{e}_{xA} + g(\mathrm{t}).\mathbf{e}_{yA}]\otimes[g(\mathrm{t}).\mathbf{e}_{xB} - f(\mathrm{t}).\mathbf{e}_{yB}] \qquad \text{(I.1a)}$$

$$\boldsymbol{\psi_{AB}}^{+}(\mathrm{t}) = [f(\mathrm{t}).\mathbf{e}_{xA} + g(\mathrm{t}).\mathbf{e}_{yA}]\otimes[g(\mathrm{t}).\mathbf{e}_{xB} + f(\mathrm{t}).\mathbf{e}_{yB}] \qquad \text{(I.1b)}$$

$$\boldsymbol{\varphi_{AB}}^{+}(\mathrm{t}) = [f(\mathrm{t}).\mathbf{e}_{xA} + g(\mathrm{t}).\mathbf{e}_{yA}]\otimes[f(\mathrm{t}).\mathbf{e}_{xB} + g(\mathrm{t}).\mathbf{e}_{yB}] \qquad \text{(I.1c)}$$

$$\boldsymbol{\varphi_{AB}}^{-}(\mathrm{t}) = [f(\mathrm{t}).\mathbf{e}_{xA} + g(\mathrm{t}).\mathbf{e}_{yA}]\otimes[f(\mathrm{t}).\mathbf{e}_{xB} - g(\mathrm{t}).\mathbf{e}_{yB}] \qquad \text{(I.1d)}$$

Vector-Bell states can be normalized. The analogous of ⟨ϕ|ϕ⟩ in QM is the projection of the vector into itself, f.ex. for $\psi_{AB}^-$:

$$|\psi_{AB}^-(t)| = |\psi_{AB}^-(t).\psi_{AB}^-(t)| = |\mathbf{e}_{\psi-}.\psi_{AB}^-(t)\times\psi_{AB}^-(t)/\psi_{AB}^-(t)| = V^2(t) \qquad (I.2)$$

This result is the same for all vector-Bell states. The value of $V^2(t)$ changes in time, but its average over the complete run is:

$$\langle\phi(t)\rangle \equiv (1/Tr).\int_0^{Tr} dt.\,\phi(t) = \mathbf{e}_\phi.(1/Tr).\int_0^{Tr} dt.\,V^2(t) = \mathbf{e}_\phi.N.u/Tr = \mathbf{e}_\phi.u/T \qquad (I.3)$$

where ϕ(t) indicates any of the four vector-Bell states and $T=Tr/N$ is the average time in which one particle is detected. Scaling $u/T=1$, all vector-Bell states are normalized. Note that $u/T$ can be interpreted as a "single particle average power", so that this scaling is physically reasonable.

Vector-Bell states are orthogonal among them. F.ex., the analogous of $\langle\psi_{AB}^-|\varphi_{AB}^+\rangle$ here is the projection (see eqs.I.1):

$$\psi_{AB}^-(t).\varphi_{AB}^+(t) = \{\psi_{AB}^-(t)\times\varphi_{AB}^+(t)\,/V^2\}.\,\mathbf{e}_{\psi-} = 0 \qquad (I.4)$$

which holds for all time values. The same is valid for $\psi_{AB}^+(t).\varphi_{AB}^-(t)$. For $\psi_{AB}^+(t).\varphi_{AB}^+(t)$ instead:

$$\psi_{AB}^+(t).\varphi_{AB}^+(t) = \mathbf{e}_{\psi+}.2.[f^2(t)+g^2(t)].\,f(t).g(t)/V^2(t) = \mathbf{e}_{\psi+}.2.V^2(t).cos[\nu(t)].sin[\nu(t)] \qquad (I.5)$$

which, at a given time, is different from zero in general. Yet, the trigonometric functions average to zero (recall V(t) and ν(t) are assumed to be independent):

$$\langle\psi_{AB}^+(t).\varphi_{AB}^+(t)\rangle = \mathbf{e}_{\psi+}.2.\int_0^{Tr} dt.\,V^2(t).cos[\nu(t)].sin[\nu(t)] = 0 \qquad (I.6)$$

The same happens for $\psi_{AB}^-(t).\varphi_{AB}^-(t)$, $\psi_{AB}^+(t).\psi_{AB}^-(t)$ and $\varphi_{AB}^+(t).\varphi_{AB}^-(t)$.

In summary: in the general case, vector-Bell states are normalized and orthogonal after time averaging over the complete run of duration *Tr*. This result remarks their statistical meaning.

*II. Hong-Ou-Mandel and entanglement swapping.*

These phenomena are involved in the loophole-free observation of the violation of Bell's inequalities using matter and photons [1,2]. When two indistinguishable single photons enter a 50%-50% beam-splitter through each of its two input gates, they both leave on the same output gate ("bunching") unless they are in the Bell state $|\psi^-\rangle$. This phenomenon is used to experimentally project an undefined two photon state into $|\psi^-\rangle$, and is crucial for entanglement swapping. This effect (Hong-Ou-Mandel) can be sketched by the vector-HV model as follows:

It is assumed that the vector-HV leaving a beam-splitter at the output gates *C,D* are related with the ones at the input gates *A,B* in the same way as fields in Optics:

$$\mathbf{V}^C(t) = [\mathbf{V}^A(t) + \mathbf{V}^B(t)]/\surd 2; \qquad \mathbf{V}^D(t) = [\mathbf{V}^A(t) - \mathbf{V}^B(t)]/\surd 2 \qquad (II.1)$$

Let define:

$$m^A{}_i = \int_{\theta_i}^{\theta_i+T} dt.|\mathbf{V}^A(t)|^2 = \int_{\theta_i}^{\theta_i+T} dt.[f^2(t) + g^2(t)] \quad \text{(II.2)}$$

where $\theta_i$ is an arbitrary time value, and $T$ is the shortest time a single photon can be detected (else, the inverse of the incident field bandwidth). An analogous expression applies for $m^B{}_i$. The *single incident photon condition*: $m^A{}_i = m^B{}_i = u$ is assumed.

Time dependence is dropped in what follows. For $\varphi_{AB}{}^+$ (see eq.I.1c) $V_x{}^A = V_x{}^B = f$, $V_y{}^A = V_y{}^B = g$. Using eq.II.1, $V_x{}^C = (f+f)/\surd 2$, $V_y{}^C = (g+g)/\surd 2$, $V_x{}^D = (f-f)/\surd 2$, $V_y{}^D = (g-g)/\surd 2$, then:

$$(V^C)^2 = (V_x{}^C)^2 + (V_y{}^C)^2 = (\surd 2f)^2 + (\surd 2g)^2 = 2.V^2(t) \quad \text{(II.3a)}$$

$$(V^D)^2 = (V_x{}^D)^2 + (V_y{}^D)^2 = [(f-f)/\surd 2]^2 + [(g-g)/\surd 2]^2 = 0 \quad \text{(II.3b)}$$

using eq.II.2 to integrate in time within each short interval, then:

$$m^C{}_i = 2.u \; ; \; m^D{}_i = 0 \quad \text{(II.4)}$$

this means that two photons are detected at one output gate and zero at the other, which is the expected bunching effect. For $\psi_{AB}{}^-$ instead (see eq.I.1a):

$$(V^C)^2 = (V_x{}^C)^2 + (V_y{}^C)^2 = [(f+g)/\surd 2]^2 + [(g-f)/\surd 2]^2 = f^2 + g^2 \quad \text{(II.5a)}$$

$$(V^D)^2 = (V_x{}^D)^2 + (V_y{}^D)^2 = [(f-g)/\surd 2]^2 + [(g+f)/\surd 2]^2 = f^2 + g^2 \quad \text{(II.5b)}$$

so that, after integrating in time within each (short) interval $[\theta_i, \theta_i+T]$:

$$m^C{}_i = m^D{}_i = u \quad \text{(II.6)}$$

therefore, single photons are detected at both output gates in the same (short) time interval, as expected. The results of the vector-HV model smoothly fit QM predictions until this point. The results for the two remaining vector-Bell states are more involved. For $\varphi_{AB}{}^-$ (see eq.I.1d):

$$(V^C)^2 = [(f+f)/\surd 2]^2 + [(g-g)/\surd 2]^2 = 2f^2 = 2.V^2.cos^2(\nu) \quad \text{(II.7a)}$$

$$(V^D)^2 = [(f-f)/\surd 2]^2 + [(g+g)/\surd 2]^2 = 2g^2 = 2.V^2.sin^2(\nu) \quad \text{(II.7b)}$$

so that, after integrating in time within each (short) interval:

$$m^C{}_i = 2\int_{\theta_i}^{\theta_i+T} dt.V^2(t) - 2\int_{\theta_i}^{\theta_i+T} dt.V^2(t).sin^2[\nu(t)] = 2.u - m^D{}_i \quad \text{(II.8a)}$$

$$m^D{}_i = 2\int_{\theta_i}^{\theta_i+T} dt.V^2(t).sin^2[\nu(t)] \neq 0 \text{ in general} \quad \text{(II.8b)}$$

The same final result is obtained for $\psi_{AB}{}^+$. Leaving aside the case $m^D{}_i = u$ (which has zero measure), one photon is detected at only one of the output gates. In this way, the vector-Bell states $\varphi_{AB}{}^-$ and $\psi_{AB}{}^+$ produce "one" detection at one gate and "zero" at the other in each time interval of duration $T$. If the time integrals are extended to the long time $Tr$, then $N = Tr/T$ photons are detected at each gate, but no photons are detected simultaneously at both gates.

Note that, for all vector-Bell states in any condition:

$$m^C{}_i + m^D{}_i = 2.u \qquad \text{(II.9)}$$

so that the number of photons is conserved in the whole process.

In summary: the only incoming vector-Bell state that produces simultaneous detections at the output gates is $\psi_{AB}^-$, in agreement with QM predictions and experimental needs.

It is pertinent to note that the situation for $\varphi_{AB}^-$ and $\psi_{AB}^+$ is not fully satisfactory for, strictly speaking, the bunching effect means "two" and "zero" single photons detected in each gate *during each time T*. This can be solved by adding an extra instruction to the way the beam splitter operates, stating (f.ex.) that the vector-HV with the smaller value of the time integral follows the path of the one with the larger value. In this way, "two" particles are detected at one gate and "zero" at the other for $\varphi_{AB}^-$ and $\psi_{AB}^+$ too. If this extra instruction is added, QM predictions for the relationships between the *polarizations* at the two output gates [3], for all vector-Bell states, are reproduced too.

The vector-HV model describes the phenomenon of entanglement swapping in a straightforward way. This is because entanglement swapping arises from a mere identity between combinations of waves. The relevant identity among QM states is:

$$|\psi_{12}^-\rangle \otimes |\psi_{34}^-\rangle = ½.\{|\psi_{14}^+\rangle \otimes |\psi_{23}^+\rangle - |\psi_{14}^-\rangle \otimes |\psi_{23}^-\rangle - |\varphi_{14}^+\rangle \otimes |\varphi_{23}^+\rangle + |\varphi_{14}^-\rangle \otimes |\varphi_{23}^-\rangle\} \qquad \text{(II.10)}$$

while the identity among vector-Bell states is instead (time dependence is dropped):

$$\psi_{12}^- \otimes \psi_{34}^- = ½.\{\psi_{14}^+ \otimes \psi_{23}^+ - \psi_{14}^- \otimes \psi_{23}^- + \varphi_{14}^+ \otimes \varphi_{23}^+ + \varphi_{14}^- \otimes \varphi_{23}^-\} \qquad \text{(II.11)}$$

The only difference is one term (the $\varphi^+$-s) with the opposite sign. It means a phase difference in a term that cancels out when the projection on $\psi^-$ is performed. It has no observable consequences in a single entanglement swapping process. Nevertheless, it has been recently shown that QM predictions in some entanglement swapping network scenarios can be reproduced by using complex numbers only [4]. Here the vector-Bell states are defined real for simplicity, but they can be defined complex as well. The different signs of the $\varphi^+$-s terms in eqs. II.11 and II.10 might be related with the difference between defining $f$(t) and $g$(t) real, or complex. This issue deserves further study.

Eq.II.11 is valid for all time values, not only in the average over time *Tr*. This is verified by writing the rhs of eq.II.11 using the following expressions, obtained directly from eqs.I.1:

$$\psi_{14}^+ \otimes \psi_{23}^+ = (f.\mathbf{e}_{x1}+g.\mathbf{e}_{y1}) \otimes (g.\mathbf{e}_{x4}+f.\mathbf{e}_{y4}) \otimes (g.\mathbf{e}_{x2}-f.\mathbf{e}_{y2}) \otimes (-f.\mathbf{e}_{x3}+g.\mathbf{e}_{y3}) \qquad \text{(II.12a)}$$

$$\psi_{14}^- \otimes \psi_{23}^- = (f.\mathbf{e}_{x1}+g.\mathbf{e}_{y1}) \otimes (g.\mathbf{e}_{x4}-f.\mathbf{e}_{y4}) \otimes (g.\mathbf{e}_{x2}-f.\mathbf{e}_{y2}) \otimes (-f.\mathbf{e}_{x3}-g.\mathbf{e}_{y3}) \qquad \text{(II.12b)}$$

$$\varphi_{14}^+ \otimes \varphi_{23}^+ = (f.\mathbf{e}_{x1}+g.\mathbf{e}_{y1}) \otimes (f.\mathbf{e}_{x4}+g.\mathbf{e}_{y4}) \otimes (g.\mathbf{e}_{x2}-f.\mathbf{e}_{y2}) \otimes (g.\mathbf{e}_{x3}-f.\mathbf{e}_{y3}) \qquad \text{(II.12c)}$$

$$\varphi_{14}^- \otimes \varphi_{23}^- = (f.\mathbf{e}_{x1}+g.\mathbf{e}_{y1}) \otimes (f.\mathbf{e}_{x4}-g.\mathbf{e}_{y4}) \otimes (g.\mathbf{e}_{x2}-f.\mathbf{e}_{y2}) \otimes (g.\mathbf{e}_{x3}+f.\mathbf{e}_{y3}) \qquad \text{(II.12d)}$$

Note that these expressions are univocally defined. Consider f.ex. eq.II.12a. If $\mathbf{V_1}$(t) is arbitrarily chosen to be $(f.\mathbf{e}_{x1}+g.\mathbf{e}_{y1})$, then $\mathbf{V_2}(t) = (g.\mathbf{e}_{x2}-f.\mathbf{e}_{y2})$ because vectors 1 and 2 are emitted in the same physical process (say, frequency down conversion in a type-II crystal) which produces a $\psi^-$ vector-state. As the form in eq.II.12a is a $\psi_{14}^+$ state, then $\mathbf{V_4}(t) = (g.\mathbf{e}_{x4}+f.\mathbf{e}_{y4})$, and $\mathbf{V_3}(t) = (-f.\mathbf{e}_{x3}+g.\mathbf{e}_{y3})$ to get a $\psi_{23}^{+}$ state. This expression for $\mathbf{V}_3(t)$ is also the one necessary to fit the $\psi_{34}^{-}$ state produced in the actual physical process of emission in the crystal. The same applies to the other three equations. Finally, the lhs in eq.II.11 is:

$$\psi_{12}^{-}\otimes\psi_{34}^{-} = (f.\mathbf{e}_{x1}+g.\mathbf{e}_{y1})\otimes(g.\mathbf{e}_{x2}-f.\mathbf{e}_{y2})\otimes(f.\mathbf{e}_{x3}+g.\mathbf{e}_{y3})\otimes(-g.\mathbf{e}_{x4}+f.\mathbf{e}_{y4}) \quad \text{(II.13)}$$

which fits the physical process of emission of pairs 1-2 and 3-4, as it must be.

For the sake of completeness, let review the process of entanglement swapping. Let suppose that two entangled pairs of systems are independently created: 1 and 2, and 3 and 4, with vector-HV related as the vector-Bell states $\psi_{12}^{-}$ and $\psi_{34}^{-}$. The four systems are then described by eq.II.11. Now let suppose that systems 1 and 4 are combined in a beam-splitter. If detections occur at both output gates in the time interval $[\theta_i,\theta_i+T]$, then, as it was discussed above, their incoming vector-HV were in the relationship corresponding to $\psi_{14}^{-}$. In consequence, during $[\theta_i,\theta_i+T]$ the vector-HV of the whole is given by the projection (filtering) of eq.II.11 on $\psi_{14}^{-}$. From the conditions of orthogonality derived in Section I, the only nonzero term is $\psi_{14}^{-}\otimes\psi_{23}^{-}$. Therefore, systems 2 and 3 show entanglement of the $\psi_{23}^{-}$ type when observed during $[\theta_i,\theta_i+T]$ (propagation times must be taken into account to correct the value of $\theta_i$). In summary: two systems that were independently created (and were thus uncorrelated) become entangled after vector filtering (projection).

*III. Eberhardt states.*

The loophole-free experiments performed using only photons [5,6] do not use maximally entangled states, but partially entangled ones, or Eberhardt states. It is certainly anti-intuitive that, to test QM vs Local Realism #2, a partially entangled state is better suited than a maximally entangled one. Eberhardt states have the form (|r|<1, typically |r| ≈ 0.3):

$$|\psi_E\rangle = (1+r^2)^{-½}\{|x_A,y_B\rangle + r.|y_A,x_B\rangle\} \quad \text{(III.1)}$$

which Concurrence is $2|r|/(1+r^2)$. The probability of coincidences is:

$$P_{AB}^{++}(\alpha,\beta) = (1+r^2)^{-1}.[cos(\alpha).sin(\beta) + r.sin(\alpha).cos(\beta)]^2 \quad \text{(III.2)}$$

and the probabilities of single detections are:

$$P_A^{+}(\alpha) = (1+r^2)^{-1}.[cos^2(\alpha) + r^2.sin^2(\alpha)] \quad \text{(III.3)}$$

$$P_B^{+}(\beta) = (1+r^2)^{-1}.[r^2.cos^2(\beta) + sin^2(\beta)] \quad \text{(III.4)}$$

Choosing the angle settings so that $cos(\alpha)\approx 0$ and $sin(\beta)\approx 0$, the single probabilities are $\approx r^2$. From eq.III.2 also $P_{AB}^{++}(\alpha,\beta)$, $P_{AB}^{++}(\alpha,\beta')$ and $P_{AB}^{++}(\alpha',\beta) \approx r^2$. The settings α',β' are still free to make $P_{AB}^{++}(\alpha',\beta') \approx 0$. Note that the fields at each station appear strongly polarized (for Bell's states instead, they are unpolarized) and that the analyzers are set almost crossed to these fields. In this way, $|\psi_E\rangle$ allows violating the corresponding Bell's inequality (Clauser-Horne) with detectors' efficiencies as low as 2/3 [7], what made a loophole-free test achievable using only photons. Other

Eberhardt state that has been used is:

$$|\varphi_E\rangle = (1+r^2)^{-1/2}.\{|x_A,x_B\rangle + r.|y_A,y_B\rangle\} \quad \text{(III.5)}$$

The probability of coincidences for this state is:

$$P_{AB}^{++}(\alpha,\beta) = (1+r^2)^{-1}.[cos(\alpha).cos(\beta) + r.sin(\alpha).sin(\beta)]^2 \quad \text{(III.6)}$$

The probabilities of single detections have the form of eq.III.3, the same for both A and B, due to the symmetry of the terms of $|\varphi_E\rangle$.

The expressions for the vector-HV are found by noting that, for QM states:

$$|\psi_E\rangle = k_1.|\psi^+\rangle + k_2.|\psi^-\rangle; \quad |\varphi_E\rangle = k_1.|\varphi^+\rangle + k_2.|\varphi^-\rangle \quad \text{(III.7)}$$

where:

$$k_1 = (1+r)/(2+2r^2)^{1/2}; \quad k_2 = (1-r)/(2+2r^2)^{1/2} \quad \text{(III.8)}$$

useful expressions are $(k_1+k_2)^2= 2/(1+r^2)$, $(k_1-k_2)^2= 2r^2/(1+r^2)$.

From eq.III.7 and taking into account eqs.I.1, the form of the vector-HV to reproduce the predictions of $|\varphi_E\rangle$ in station A is intuitively proposed as:

$$\mathbf{V}_A(t) = k_1.[f(t).\mathbf{e}_x + g(t).\mathbf{e}_y] + k_2.[f(t).\mathbf{e}_x - g(t).\mathbf{e}_y] \quad \text{(III.9)}$$

For, the first (second) term has the symmetry of the vector-Bell state $\varphi^+$ ($\varphi^-$) analogous to $|\varphi^+\rangle$ ($|\varphi^-\rangle$). The number $N_\alpha$ of detections after the analyzer in station A is given, as always, by the projected vector (time dependence of *f*,*g* is dropped):

$$\mathbf{e}_\alpha.\mathbf{V}_A(t) = \mathbf{e}_\alpha.k_1.[f.cos(\alpha)+g.sin(\alpha)] + \mathbf{e}_\alpha.k_2.[f.cos(\alpha)-g.sin(\alpha)] = \quad \text{(III.10)}$$

$$= \mathbf{e}_\alpha.[f.cos(\alpha).(k_1+k_2) + g.sin(\alpha).(k_1-k_2)]$$

therefore:

$$N_\alpha = (1/u).\int_0^{Tr} dt.\, |\mathbf{e}_\alpha.\mathbf{V}_A(t)|^2 = (1/u).\int_0^{Tr} dt.\, \{ f^2.cos^2(\alpha).(k_1+k_2)^2 + 2.f.g.cos(\alpha).sin(\alpha).(k_1+k_2).(k_1-k_2) +$$

$$+ g^2.sin(\alpha).(k_1-k_2)^2\} \quad \text{(III.11)}$$

Because, from eq.7 in the main text and from eq.I.5 here, the integrals in $f^2$ and $g^2$ are ½ *N* and the one in *f.g* is zero. Then:

$$N_\alpha = ½.N.\{cos^2(\alpha).(k_1+k_2)^2 + sin(\alpha).(k_1-k_2)^2\} = [N/(1+r^2)].\{cos^2(\alpha) + r^2.sin(\alpha)\} \quad \text{(III.12)}$$

which is equivalent to eq.III.3. Thus, the proposed form for $\mathbf{V}_A$ is correct. The same approach applies to $N_\beta$ in the space of station B, or B-space.

The number of coincidences is given by the projection of the corresponding component (of $\mathbf{e}_\alpha.\mathbf{V}_A$, translated into B-space), named $\mathbf{V}_{corr}$, into $\mathbf{e}_\beta$. The expression of $\mathbf{V}_{corr}$ is found by transforming the angles to the B-space as indicated in eq.14 in the main text. That is: for the symmetry of vector state $\varphi^+$ ($\varphi^-$), $\nu_B = \nu_A$ ($\nu_B = -\nu_A$). Then eq.III.10 is written in the B-space as:

$$\mathbf{V}_{corr} = \mathbf{e}_\alpha.k_1.[f.cos(\alpha)+g.sin(\alpha)] + \mathbf{e}_{-\alpha}.k_2.[f.cos(\alpha)+g.sin(\alpha)] \quad \text{(III.13)}$$

Recall that the term multiplied by $k_1$ ($k_2$) corresponds to $\varphi^+$ ($\varphi^-$), be aware of the ($-\alpha$) sub index in the unit vector in the second term. Then:

$$|\mathbf{e}_\beta.\mathbf{V}_{corr}|^2 = \{cos(\alpha-\beta).k_1 + cos(\alpha+\beta).k_2\}^2.[f.cos(\alpha)+g.sin(\alpha)]^2 \quad \text{(III.14)}$$

After time integration over *Tr* and some algebra of trigonometric functions:

$$N^{++} = [N/(1+r^2)].\{cos(\alpha).cos(\beta) + r.sin(\alpha).sin(\beta)\}^2 \quad \text{(III.15)}$$

which is the correct result, see eq.III.6.

The predictions for $|\psi_E\rangle$ are obtained in the same way. For station A, now using $\psi^+$ and $\psi^-$:

$$\mathbf{V}_A(t) = k_1.[g.\mathbf{e}_x + f.\mathbf{e}_y] + k_2.[g.\mathbf{e}_x - f.\mathbf{e}_y] \quad \text{(III.16)}$$

To calculate the vector-HV in the B-space, take into account that for $\psi^+$ ($\psi^-$), $\nu_B = \nu_A-\pi/2$ ($\nu_B = \pi/2-\nu_A$). In consequence, the angle of the component coming from $\psi^+$ is shifted a constant angle $\pi$ with respect to the component coming from $\psi^-$. This means a factor (-1) in the second term of eq.III.16 when translated into the B-space, hence:

$$\mathbf{V}_B(t) = k_1.[g.\mathbf{e}_x + f.\mathbf{e}_y] + k_2.[-g.\mathbf{e}_x + f.\mathbf{e}_y] \quad \text{(III.17)}$$

This leads, after projection on $\mathbf{e}_\beta$ and time integration over *Tr*, to eq.III.4.

The number of coincidences is found as in the $|\varphi_E\rangle$ case, taking into account the angle transformations $\nu_B = \nu_A-\pi/2$ ($\nu_B = \pi/2-\nu_A$) for the $\psi^+$ ($\psi^-$) term, and the (-1) factor just explained. From eq.III.16, and as in eq.III.13, then:

$$\mathbf{V}_{corr} = \mathbf{e}_{(\pi/2-\alpha)}.k_1.[g.cos(\pi/2-\alpha)+f.sin(\pi/2-\alpha)] + (-1).\mathbf{e}_{(-\pi/2+\alpha)}.k_2.[g.cos(-\pi/2+\alpha)-f.sin(-\pi/2+\alpha)] \quad \text{(III.18)}$$

so that:

$$|\mathbf{e}_\beta.\mathbf{V}_{corr}|^2 = [g.sin(\alpha)+f.cos(\alpha)]^2.\{k_1.[sin(\alpha).cos(\beta)+cos(\alpha).sin(\beta)] - - k_2.[sin(\alpha).cos(\beta)-cos(\alpha).sin(\beta)]\}^2 \quad \text{(III.19)}$$

After time integration over *Tr* it leads to eq.III.2, which is the correct result.

## References of the Supplementary Material Section.